\documentclass[aps,prl,showpacs,notitlepage,amsmath,amssymb,floatfix,twocolumn,superscriptaddress]{revtex4-2}

\usepackage[utf8]{inputenc}
\usepackage[T1]{fontenc}
\usepackage[english]{babel}

\usepackage{amsmath}
\usepackage{amssymb}
\usepackage{siunitx}
\usepackage{pgfplots}
\pgfplotsset{compat=1.18}
\usetikzlibrary{arrows.meta}
\usepackage{tikz}
\usepackage{pgfplots}
\usepackage{tikz}
\usepackage{pgffor}

\usepackage{graphicx}
\usepackage[svgnames]{xcolor}
\usepackage{geometry}
\usepackage{fancyhdr}
\usepackage{titlesec}
\usepackage{booktabs}
\usepackage{caption}
\usepackage{subcaption}
\usepackage{enumitem}
\setlist{nosep}
\usepackage{tcolorbox}
\tcbuselibrary{breakable}
\usepackage{multirow}
\usepackage{float}

\usepackage{amsmath}
\usepackage{amssymb}
\usepackage{hyperref}
\usepackage[normalem]{ulem}
\usepackage{mathtools}
\usepackage{comment}

\DeclareMathAlphabet{\altmathcal}{OMS}{cmsy}{m}{n}
\usepackage{mathptmx}

\usepackage{booktabs}
\usepackage{xcolor}
\usepackage{amsmath}
\usepackage{graphicx} % Required for inserting images
\usepackage{subcaption}
\usepackage{soul}
\usepackage{babel}
\usepackage{ulem}

\newcommand{\NEW}[1]{#1}
\hypersetup{colorlinks, allcolors=black}

\begin{document}

\title{Dynamical hysteresis in the dissipation in turbulent flows}

%%xxxxxxxxxxxxxxxxxxxxxxxxxxxxxxxxxxxxxxxxxxxxxxxxxxxxxxxxxxxxxxxxxxxxxxxxxxxxxxxxxxxxxxxxxxxxxxxxxxxxxxxxxxxxxxxxxxx
%%xxxxxxxxxxxxxxxxxxxxxxxxxxxxxxxxxxxxxxxxxxxxxxxxxxxxxxxxxxxxxxxxxxxxxxxxxxxxxxxxxxxxxxxxxxxxxxxxxxxxxxxxxxxxxxxxxxx
%%xxxxxxxxxxxxxxxxxxxxxxxxxxxxxxxxxxxxxxxxxxxxxxxxxxxxxxxxxxxxxxxxxxxxxxxxxxxxxxxxxxxxxxxxxxxxxxxxxxxxxxxxxxxxxxxxxxx
\author{M. Ahmad}
%\email[Corresponding author: ]{mohammad.ahmad@icam.fr}

\affiliation{Univ. Lille, CNRS, ONERA, Arts et Metiers Institute of Technology, Centrale Lille, UMR 9014-LMFL-Laboratoire de M\'ecanique des Fluides de Lille - Kamp\'e de F\'eriet, F-59000 Lille, France}
%\affiliation{Icam School of Engineering, Lille campus, 6 rue Auber, B.P. 10079, 59016, Lille, France}

\author{P.D. Mininni}
\email[Corresponding author: ]{mininni@df.uba.ar}
\affiliation{Universidad de Buenos Aires, Facultad de Ciencias Exactas y Naturales,
Departamento de Física, Ciudad Universitaria, 1428 Buenos Aires, Argentina}
\affiliation{CONICET - Universidad de Buenos Aires, Instituto de Física Interdisciplinaria y Aplicada (INFINA),
Ciudad Universitaria, 1428 Buenos Aires, Argentina}

\author{M. Obligado}
\email[Corresponding author: ]{martin.obligado@centralelille.fr}
\affiliation{Univ. Lille, CNRS, ONERA, Arts et Metiers Institute of Technology, Centrale Lille, UMR 9014-LMFL-Laboratoire de M\'ecanique des Fluides de Lille - Kamp\'e de F\'eriet, F-59000 Lille, France}
\affiliation{Institut universitaire de France (IUF), Paris, France}

\author{J.A. Farnsworth}
\email[Corresponding author: ]{john.farnsworth@colorado.edu}
\affiliation{Ann and H.J. Smead Department of Aerospace Engineering Sciences, University of Colorado Boulder, Boulder, CO, USA}

%%xxxxxxxxxxxxxxxxxxxxxxxxxxxxxxxxxxxxxxxxxxxxxxxxxxxxxxxxxxxxxxxxxxxxxxxxxxxxxxxxxxxxxxxxxxxxxxxxxxxxxxxxxxxxxxxxxxx
%%xxxxxxxxxxxxxxxxxxxxxxxxxxxxxxxxxxxxxxxxxxxxxxxxxxxxxxxxxxxxxxxxxxxxxxxxxxxxxxxxxxxxxxxxxxxxxxxxxxxxxxxxxxxxxxxxxxx

\begin{abstract}
We present evidence of the dynamical hysteretic nature of dissipation in unsteady turbulent flows. Wind tunnel experiments and direct numerical simulations in oscillating flows show that, at stationary mean Reynolds number, the dissipation constant is larger for decelerating flows. Consequently, a periodic behavior of the flow produces a hysteresis cycle, whose area scales with a parameter combining the Strouhal number and the relative amplitude of the forcing. This phenomenon can be explained and quantified through the influence of the unsteady term in the Kármán–Howarth equation, with implications for a wide range of out-of-equilibrium systems.
\end{abstract}

        \maketitle
%%xxxxxxxxxxxxxxxxxxxxxxxxxxxxxxxxxxxxxxxxxxxxxxxxxxxxxxxxxxxxxxxxxxxxxxxxxxxxxxxxxxxxxxxxxxxxxxxxxxxxxxxxxxxxxxxxxxx
%%xxxxxxxxxxxxxxxxxxxxxxxxxxxxxxxxxxxxxxxxxxxxxxxxxxxxxxxxxxxxxxxxxxxxxxxxxxxxxxxxxxxxxxxxxxxxxxxxxxxxxxxxxxxxxxxxxxx
%%xxxxxxxxxxxxxxxxxxxxxxxxxxxxxxxxxxxxxxxxxxxxxxxxxxxxxxxxxxxxxxxxxxxxxxxxxxxxxxxxxxxxxxxxxxxxxxxxxxxxxxxxxxxxxxxxxxx
%\section{Introduction}
Hysteresis arises whenever a system retains memory of its past states. Classical hysteresis is associated with bistability~\cite{chow2023}: the system settles on different branches depending on its history, as in ferromagnetic magnetisation or laminar-to-turbulent transition. A distinct class is dynamical hysteresis~\cite{broner1997}, in which a periodically driven system traces a loop whose area vanishes as the forcing frequency tends to zero, reflecting a finite relaxation time rather than multistability. Observing such behavior in turbulence---a system far from equilibrium, traditionally regarded as rapidly mixing and weakly history-dependent---would reveal a previously unrecognized form of memory in the energy cascade, with broad implications for driven dissipative systems.

Turbulent flows are rarely stationary: atmospheric and oceanic flows, industrial mixers, and engineering devices are all subjected to time-dependent forcing whose periodic or quasi-periodic character spans a wide range of timescales, from \NEW{atmospheric gusts} to industrial pulsations and \NEW{flapping wings}. A natural framework is provided by the K\'arm\'an--Howarth (KH) equation, which describes the scale-by-scale energy balance in homogeneous isotropic turbulence and explicitly incorporates large-scale temporal variations:
\begin{equation}
%\begin{multline}
    % \underbrace{-\frac{\overline{\delta u^{3}}}{r}}_{\text{(I) Inertial Flux}} + \underbrace{\frac{6\nu}{r}\frac{\partial}{\partial r}\overline{\delta u^{2}}}_{\text{(II) Viscous Diffusion}}
    % + \underbrace{\frac{3}{r^{5}}\int_{0}^{r}s^{4}\frac{\partial}{\partial t}\overline{\delta u^{2}}ds}_{\text{(III) Unsteady Term}} \\ = \underbrace{\frac{4}{5}\overline{\epsilon}}_{\text{(IV) Dissipation}}.
    -\frac{\overline{\delta u^{3}}}{r} + \frac{6\nu}{r}\frac{\partial}{\partial r}\overline{\delta u^{2}}
    + \frac{3}{r^{5}}\int_{0}^{r}s^{4}\frac{\partial}{\partial t}\overline{\delta u^{2}}ds \\ = \frac{4}{5}{\epsilon},
    \label{eq:karman_howarth_main}
%\end{multline}
\end{equation}
that can be exactly derived from the Navier-Stokes equations under the assumptions of homogeneity and isotropy~\cite{monin2013statistical}. In this equation, $\nu$ is the kinematic viscosity of the fluid, $\varepsilon$ is the turbulent energy dissipation rate, $r$ is the spatial increment in the second and third order averaged structure functions ($\overline{\delta u^{2}}$ and $\overline{\delta u^{3}}$, respectively), and $s$ is a dummy integration variable in the unsteady term. The first term in the left-hand side is related to the inter-scale energy transfer, the second to viscous effects, and the third to large-scale unsteadiness. The term on the right hand side quantifies the turbulent dissipation across scales.

Within the Kolmogorov phenomenology, the turbulent inertial range corresponds to scales $r$ that are large (resp.~small) enough that the viscous term (resp.~unsteady term) can be neglected. In consequence, the well-known $4/5$ law is obtained, that in its integral form implies that the dimensionless dissipation, $C_{\epsilon} = \epsilon L / u'^3$ (with $L$ the integral scale of turbulence and $u'$ the rms value of velocity fluctuations), is constant at large Reynolds numbers for a given boundary condition~\cite{pope2000turbulent,sreenivasan1995universality}. However, a significant body of modern research has revealed that many canonical turbulent flows, in certain regimes, do not fulfill the Kolmogorov balance but still present an inertial range, where $C_{\epsilon}$ is not constant but scales with the local Taylor-scale Reynolds number, $Re_{\lambda}$~\cite{vassilicos2015dissipation}. This departure from classical theory is, among other causes, a direct consequence of the unsteady nature of the nonlinear energy transfer across scales.

This work focuses on flows with deliberate unsteady forcing, containing frequencies that can contaminate the inertial range of turbulence. In all cases, the forcing consists of a solenoidal function that simplifies modeling via Eq.~(\ref{eq:karman_howarth_main}). For such an aim, a wind tunnel with an active system was used to produce a pulsed flow with frequencies of up to 1 Hz \cite{farnsworth2020design}. One-dimensional hot-wire anemometry (HWA) measurements were carried out to resolve all relevant scales. The combined use of a phase-averaging, triple decomposition, and Taylor's hypothesis allowed for the quantification of all the terms in the KH equation. To validate our findings without these assumptions, direct numerical simulations (DNS) were also carried out. As detailed below, the numerical results agree closely with the experiments, while resolving all terms in Eq.~(\ref{eq:karman_howarth_main}) without requiring any decomposition or additional hypotheses.

Although several studies have addressed the role of large-scale forcing in turbulent flows \cite{goto2016energy,steiros2022balanced,bos2017dissipation}, little attention has been paid to the emergence of hysteresis in the energy cascade. The most direct physical manifestation of a non-zero unsteady term with a timescale close to the inertial range is the development of a memory within the latter \cite{goto2016energy,obligado2019quantifying,meldi2021analysis}. When the flow is forced periodically, this memory is expected to reveal itself as a hysteresis loop in the phase-averaged ($\langle C_\epsilon \rangle_\phi$, $\langle Re_\lambda \rangle_\phi$) plane. The system would hence follow one path during the accelerating phase of the forcing, and a different path during the decelerating phase, demonstrating that the dissipative state of turbulence depends not just on the instantaneous Reynolds number, but on its recent history. While traces of its possible existence have been recently reported \cite{zapata2024turbulence,bos2025equilibrium}, a fundamental problem remains, which constitutes the main goal of this work: there is no established quantitative framework that links the macroscopic parameters of the external forcing, its frequency and amplitude, to the resulting magnitude of the hysteresis.

\begin{figure}
    \centering
    % Subfigure (a) for Triple Decomposition
    \begin{subfigure}[b]{\linewidth}
        \centering
        \includegraphics[width=0.95\linewidth, clip]{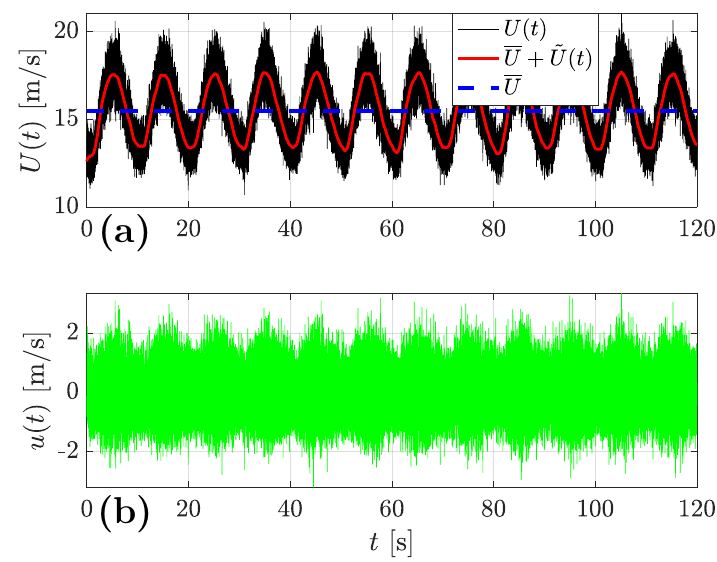}%, trim=1cm 7.4cm 2cm 8.2cm%trim=2cm 9cm 2cm 9cm
        % \fbox{\parbox[c][6.5cm][c]{0.9\linewidth}{\centering \textbf{Placeholder for Figure 1a: Triple Decomposition} }}
        % \caption{}
        \label{fig:triple_decomposition}
    \end{subfigure}
    
    \vspace{0.5cm}
    
    % Subfigure (b) for Hysteresis Loop
    \begin{subfigure}[b]{\linewidth}
        \centering
        \includegraphics[width=0.95\linewidth, clip,]{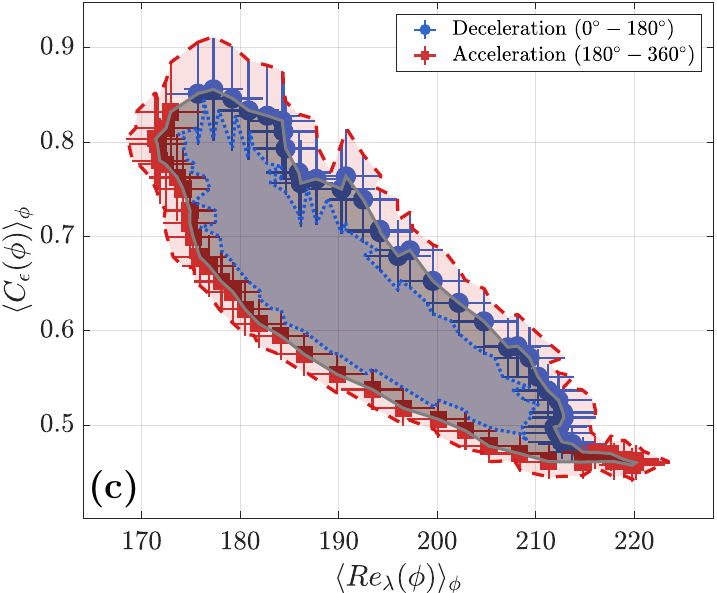}% trim=1.5cm 7.4cm 2.7cm 8.4cm
        % \includegraphics[width=0.95\linewidth, clip, trim=2.8cm 9.1cm 3.5cm 9.7cm]{./Smooth_Loop_Plot_FS_F_1Hz_A_0_20.pdf}
        % \fbox{\parbox[c][6.5cm][c]{0.9\linewidth}{\centering \textbf{Placeholder for Figure 1b: Hysteresis Loop} }}
        % \caption{}
        \label{fig:hysteresis_loop_example}
    \end{subfigure}
    
    \caption{(a) Example of the triple decomposition of an experimental velocity signal (black) into its mean (blue), periodic (red), and (b) turbulent (green) components. (c) Phase- and ensemble-averaged hysteresis loop (grey solid polygon) in the ($\langle Re_{\lambda}\rangle_{\phi}$, $\langle C_{\epsilon}\rangle_{\phi}$) plane for a representative case, showing the distinct decelerating (blue circles) and accelerating (red squares) paths; error bars represent one standard deviation from both the phase- and ensemble-averaging. The outer (red dashed) and inner (blue dotted) shaded regions delimit the maximum and minimum hysteresis extents.
}
% (b) The resulting phase-averaged hysteresis loop in the ($\langle Re_{\lambda} \rangle_{\phi}$, $\langle C_{\epsilon} \rangle_{\phi}$) plane for a representative case, showing the distinct decelerating (blue) and accelerating (red) paths. The shaded area in magenta represents the area of the hysteresis.
%, $\Delta A_H$. $\varepsilon$ and the Taylor scale $\lambda$ were estimated using standard spectral techniques~\cite{mora2020estimating}
    \label{fig:concepts}
\end{figure}

The experiments were conducted in a low-speed wind tunnel which could easily house different passive turbulence-generating grids, two of which were used for this study: a four-generation fractal square grid (FS, first-generation length $M=195$~mm), and a large classical grid (LG, mesh size $M=102$~mm). The freestream velocity was dynamically modulated via a set of louvers situated far upstream, at the inlet to the wind tunnel fan, producing a precisely controlled sinusoidal forcing~\cite{farnsworth2020design}. The flow was characterized using a single-component HWA system positioned $3$~m downstream of the grids. Ten repeated time series of $120$~s were acquired at a 50 kHz sampling frequency for each case to resolve all relevant turbulent scales. We employ the triple decomposition $U(t) = \overline{U} + \widetilde{U}(t) + u(t)$ \cite{hussain1970mechanics} to separate the coherent periodic component $\widetilde{U}$ (extracted by a narrow-band IIR filter) from the turbulent fluctuations $u$; the full extraction procedure is described in \hyperref[em:decomp]{the End Matter (S2)}. An example is shown in Fig.~\ref{fig:concepts}(a). Phase-binning of the turbulence signals into $20^\circ$ bins sorted by the forcing phase yields phase-averaged statistics within each bin~\cite{zheng2023unsteady}. For the FS grid, we explored $f = 0.05$--$1$~Hz at fixed $|\widetilde{U}|/\overline{U} \approx 0.2$, and varied $|\widetilde{U}|/\overline{U}$ from $0.05$ to $0.20$ at $f = 0.5$~Hz. For the LG grid, $f = 0.1$, $0.5$, and $1$~Hz were investigated at $|\widetilde{U}|/\overline{U} \approx 0.2$. The mean velocity $\overline{U} \approx 15$~m/s yielded $\mathrm{Re}_\lambda \approx 196$ (FS) and $230$ (LG); a complete parameter table with all conditions and $X$ values is given in \hyperref[tab:S1]{the End Matter (Table~I)}.

The primary signature of the unsteady energy cascade is the resulting hysteresis loop observed in the ($\langle Re_{\lambda} \rangle_{\phi}$, $\langle C_{\epsilon} \rangle_{\phi}$) plane, shown in Fig.~\ref{fig:concepts}(b), where the system follows different paths during the accelerating and decelerating phases. We characterize the magnitude of the hysteresis by its normalized area $\Delta \langle C_{\epsilon}(\phi) \rangle_\phi$, which is calculated by integrating the area enclosed by the decelerating and accelerating branches of the loop $\Delta A_H$ and normalizing it by the area of its bounding box $\Delta A_N$. The goal is to establish the link between the external forcing and this observable $\langle (Re_{\lambda} \rangle_{\phi}$, $\langle C_{\epsilon} \rangle_{\phi}$) hysteresis.

These extensive experimental results are complemented by DNSs, which provide full spatio-temporal data for validation, albeit at the cost of providing less
statistics over time, with significantly less forcing cycles per DNS when compared with the experiments. The incompressible Navier-Stokes equations are solved in a three-dimensional $2\pi L_0$-periodic cubic box (with a spatial resolution of $N^3=512^3$ grid points) with a parallel pseudo-spectral method using the GHOST code \cite{mininni2011, rosemberg2020}. The kinematic viscosity is $\nu_{512} = 5 \times 10^{-4}~L_0 \overline{U}$. The external forcing is given by a superposition of all modes in the $kL_0 \in [1,2)$ Fourier shell, generated with random phases. The flow is forced with constant forcing amplitude $A$ until reaching a turbulent steady state. Afterwards, the global amplitude of the forcing is modulated with amplitude $A + \delta A \sin (2 \pi t/T)$, integrating the flow for several cycles. Simulations were performed with $\delta A/A = 1/4$ and $T=50 L_0 / \overline{U}$, $\delta A/A = 1/4$ and $T=25 L_0 / \overline{U}$, and $\delta A/A = 1/8$ and $T=50 L_0 / \overline{U}$. Spatial isotropic energy spectra, and the full velocity field ${\bf u}({\bf x},t)$, allow direct computation of all quantities below without assuming Taylor's hypothesis. As for the HWA, the values of $\varepsilon$ and $\lambda$ were estimated according with standard techniques that are consistent with the HWA calculations~\cite{zapata2024turbulence}. The KH equation balance was verified for both HWA and DNS (\hyperref[fig:S3]{End Matter, Fig.~4).}%K\'arm\'an--Howarth

The hysteresis phenomenon introduced in Fig.~\ref{fig:concepts}(b) is not an isolated case: a gallery shown in \hyperref[fig:S1]{the End Matter (Fig.~5)} confirms that well-defined loops are obtained for all experimental and DNS configurations, with size and shape depending systematically on the forcing parameters.

% \begin{figure}
%     \centering
%     % Subfigure (a) for Experimental Hysteresis Loops
%     \begin{subfigure}[b]{0.49\textwidth}
%         \centering
%         \includegraphics[width=0.95\linewidth, clip, trim=0.85cm 7.6cm 2.0cm 7.9cm]{../The_onset_of_hysteresis_in_the_energy_cascade/./Gallery_Hysteresis_Exp_FS_LG.pdf}
%         % \fbox{\parbox[c][6cm][c]{\linewidth}{\centering \textbf{Placeholder for Figure 2a}}}
%         % \caption{}%Experimental (HWA) Results
%         \label{fig:hwa_loops}
%     \end{subfigure}
%     \hfill % space between subfigures
%     % Subfigure (b) for Numerical Hysteresis Loops
%     \begin{subfigure}[b]{0.49\textwidth}
%         \centering
%         \includegraphics[width=0.95\linewidth, clip, trim=0.85cm 7.6cm 2.0cm 7.9cm]{../The_onset_of_hysteresis_in_the_energy_cascade/./Gallery_Hysteresis_DNS.pdf}%, clip, trim=1cm 7.4cm 2cm 8.2cm%trim=2.8cm 9.1cm 3.5cm 9.7cm
%         % \caption{Numerical (DNS) Results}
%         \label{fig:dns_loops}
%     \end{subfigure}

%     \caption{Gallery of phase-averaged hysteresis loops in the ($\langle Re_{\lambda} \rangle_{\phi}$, $\langle C_{\epsilon} \rangle_{\phi}$) plane for various forcing conditions, demonstrating the robustness of the phenomenon in both (a) experiments and (b) simulations.}
%     \label{fig:multiple_hysteresis}
% \end{figure}

To move from a qualitative observation to a predictive framework, we now focus on the relationship between the forcing and the resulting hysteresis. By inspecting the unsteady term in Eq.~(\ref{eq:karman_howarth_main}), we propose a new dimensionless parameter, $X = St_f \cdot (|\widetilde{U}|/\overline{U})$, that combines the Strouhal number based on the forcing frequency, $St_f = (L / u') \cdot f$, where $L$ is the integral length scale and $u'$ is the RMS of the velocity fluctuations, and the relative amplitude of the velocity modulation, $|\widetilde{U}|/\overline{U}$. This parameter represents a complete measure of the large-scale unsteadiness imposed on the flow. For the DNS, the role of $|\widetilde{U}|/\overline{U}$ is played by $\delta A/A$, the forcing amplitude modulation ratio, which is the analogous control parameter. The necessity of including the amplitude factor (rather than $\mathrm{St}_f$ alone) is demonstrated in \hyperref[fig:S4]{the End Matter (Fig.~7)}. In experiments, $L$ is calculated, following a standard procedure, integrating the autocorrelation function $R_u$ until the first zero crossing~\cite{pope2000turbulent}. On the other hand, for the DNS it is estimated via the energy spectrum $E(\kappa)$, as $L=\pi/(2u'^2)\int E(\kappa)/\kappa d\kappa$, where $\kappa$ is the wavenumber in Fourier space. We remark that both formulae are mathematically equivalent.

\begin{figure}
    \centering
    \includegraphics[width=0.95\linewidth, clip, trim=1.2cm 7.4cm 2.7cm 8.4cm]{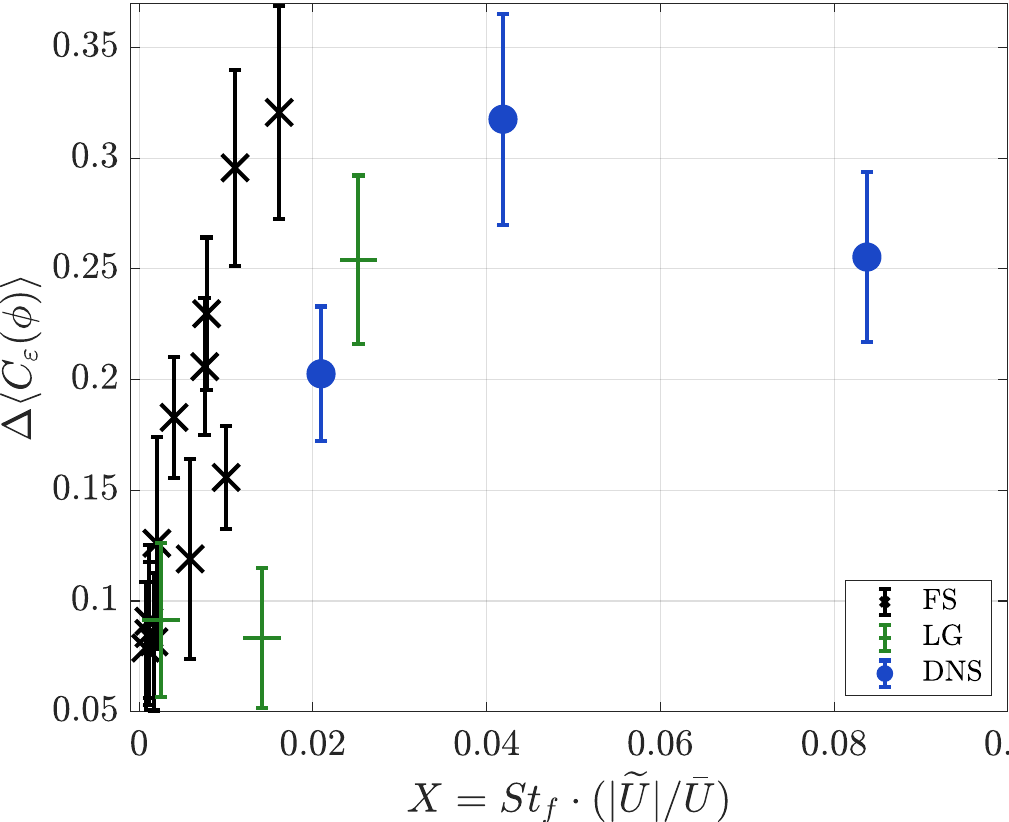}%trim=0.85cm 7.6cm 2.0cm 7.9cm
    % \fbox{\parbox[c][6cm][c]{0.49\textwidth}{\centering \textbf{Placeholder for Figure 3}}}
    
    \caption{Normalized hysteresis area, $\Delta \langle C_{\epsilon}(\phi) \rangle_\phi$, as a function of the dimensionless forcing parameter $X = St_f \cdot (|\widetilde{U}|/\overline{U})$. The data includes all experimental (HWA) and numerical (DNS) configurations.}% The dashed line represents a linear fit to the data.
    \label{fig:main_scaling_law}
\end{figure}

In particular, we find that hysteresis amplitude and the dimensionless number are correlated. Fig.~\ref{fig:main_scaling_law} shows the normalized hysteresis area, $\Delta \langle C_{\epsilon}(\phi) \rangle_\phi$, as a function of the dimensionless forcing parameter $X$ for all our experimental and numerical cases. It can be observed that $\Delta \langle C_{\epsilon}(\phi) \rangle_\phi$ increases with $X$, with a behavior that depends on the inflow conditions. For instance, in the FS grid case, which corresponds to the configuration with the largest number of data points, a linear relationship between the two parameters is observed. Although only three points are available for the LG grid and the DNS, a similar trend can be identified, albeit with a larger scatter. Despite the relevance of these observations, the estimation of $\Delta \langle C_{\epsilon}(\phi) \rangle_\phi$ carries uncertainties, as indicated by the nontrivial shapes of the hysteresis cycles shown in \hyperref[fig:S1]{Fig.~5 of the End Matter}.

%\MA{Remarkably, the data from each configuration can, despite some scatter, be approximated by a linear trend. Across the different configurations, a linear dependence can still be observed, albeit without a complete collapse. This finding provides a predictive framework that quantitatively links the macroscopic forcing parameters of an unsteady turbulent flow to the magnitude of the resulting hysteresis in its energy cascade.}

% \begin{figure}
%     \centering
%     \includegraphics[width=0.95\linewidth, clip, trim=0.85cm 7.6cm 2.0cm 7.9cm]{../The_onset_of_hysteresis_in_the_energy_cascade/./Hysteresis_vs_St_Amp_Custom_NEW_CorrectX_withoutFIT_no-fits_0crossing.pdf}
%     % \fbox{\parbox[c][6cm][c]{0.49\textwidth}{\centering \textbf{Placeholder for Figure 3}}}

%     \caption{Normalized hysteresis area, $\Delta \langle C_{\epsilon}(\phi) \rangle_\phi$, as a function of the dimensionless forcing parameter $X = St_f \cdot (|\widetilde{U}|/\overline{U})$. The data includes all experimental (HWA) and numerical (DNS) configurations.}% The dashed line represents a linear fit to the data.
%     \label{fig:main_scaling_law}
% \end{figure}

Remarkably, the trends presented in Fig.~\ref{fig:main_scaling_law} can be directly linked to the physics of the unsteady energy cascade. Indeed, the hysteresis in the dissipative state can be quantified by the non-zero unsteady term in Eq.~(\ref{eq:karman_howarth_main}). We quantify this term using the non-dimensional unsteady function $F(r)$~\cite{obligado2019quantifying}, which is defined as,
\begin{equation}
    F(r) \equiv -\frac{3}{\epsilon r^5} \int_0^r s^4 \frac{\partial}{\partial t} \overline{\delta u^{2}} ds,
    \label{eq:unsteady_F}
\end{equation}
and represents the normalized, large-scale unsteadiness of the flow.
% proportional to the time derivative of the second-order structure function and represents the normalized, large-scale unsteadiness of the flow \cite{obligado2019quantifying}:

\begin{figure*}[t!]
    \centering
    % Subfigure (a) for Phase-Averaged F(r)
    \begin{subfigure}[b]{0.33\textwidth}
        \centering
        \includegraphics[width=0.95\linewidth, clip]{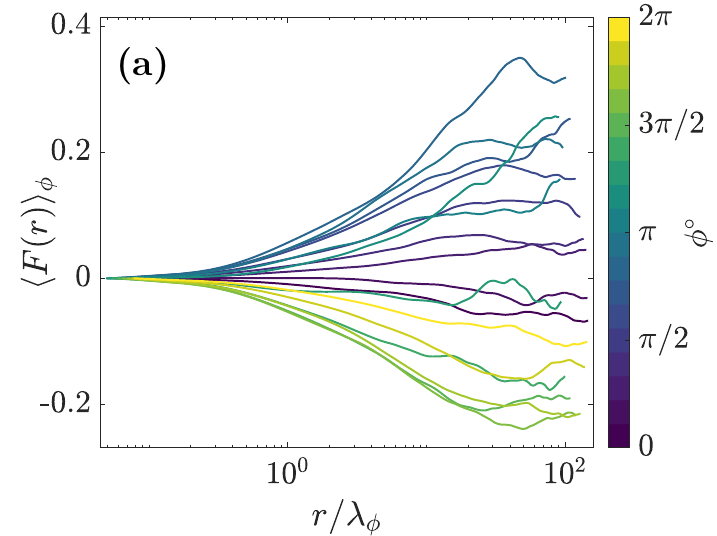}%, trim=0.65cm 7.4cm 1.2cm 8cm
        % \fbox{\parbox[c][6cm][c]{\linewidth}{\centering \textbf{Placeholder for Figure 4a}}}
        % \caption{}
        \label{fig:F_vs_phase}
    \end{subfigure}
    \hfill % space between subfigures
    \begin{subfigure}[b]{0.33\textwidth}
        \centering
        \includegraphics[width=0.95\linewidth, clip]{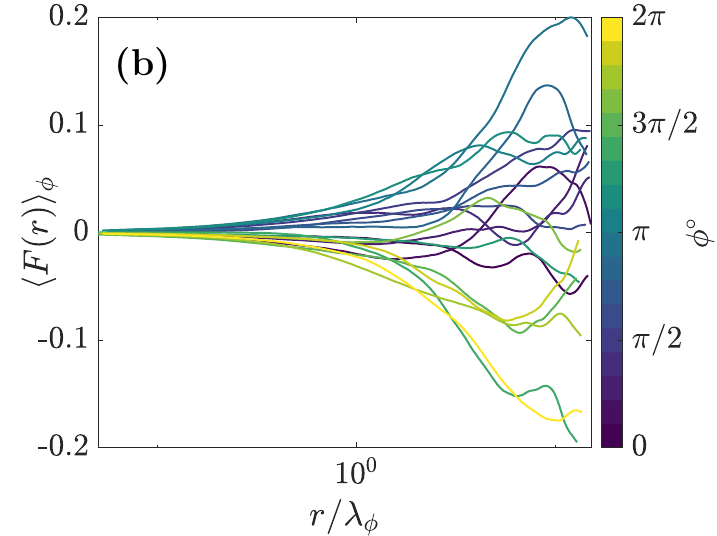}%, trim=0.9cm 7.4cm 1.2cm 8.2cm
        %\caption{}
        \label{fig:F_vs_phase_dns}
    \end{subfigure}
    \hfill % space between subfigures
    % Subfigure (b) for the F vs. Forcing Parameter Correlation
    \begin{subfigure}[b]{0.33\textwidth}
        \centering
        \includegraphics[width=0.95\linewidth, clip]{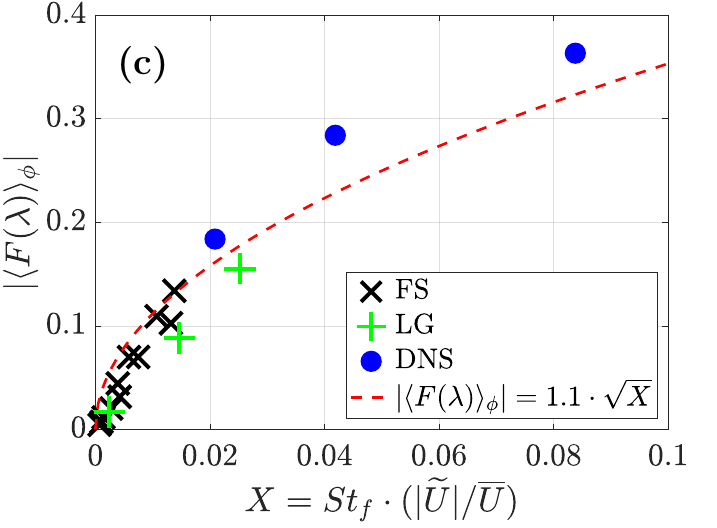}%, trim=0.65cm 7.4cm 1.2cm 8.2cm
        % \fbox{\parbox[c][6cm][c]{\linewidth}{\centering \textbf{Placeholder for Figure 4b}}}
        % \caption{}
        \label{fig:F_vs_forcing}
    \end{subfigure}
    
    \caption{(a) Experimental result of the phase-averaged profiles of the unsteady function, $\langle F(r) \rangle_{\phi}$, versus the non-dimensional lag, $r/\lambda_{\phi}$, for a representative case. (b) DNS result, for a representative case. The curves in (a) and (b) are colored by phase, where phases $\phi \in [0, \pi)$ correspond to the decelerating part of the cycle (cool colors) and $\phi \in [\pi, 2\pi)$ to the accelerating part (warm colors). (c) Scaling of the peak-to-peak amplitude of the unsteady function evaluated at the Taylor scale, $|\langle F(\lambda) \rangle_{\phi}|$, with the dimensionless forcing parameter, $X$. The scaling includes all experimental and numerical data.}
    \label{fig:unsteady_function}
\end{figure*}

Fig.~\ref{fig:unsteady_function}(a) shows the phase-averaged unsteady function, $\langle F(r) \rangle_{\phi}$, plotted against the non-dimensional lag, $r/\lambda_{\phi}$ (where $\lambda_{\phi}$ is the Taylor microscale of the flow~\cite{lundgren2002,lundgren2003}), for a representative experimental case. The function's shape and magnitude evolve significantly throughout the forcing cycle, confirming that the energy balance is in a constant state of flux. During the decelerating half of the cycle ($\phi \in [0, \pi]$), $\langle F \rangle_{\phi}$ is predominantly positive, indicating a net release of energy from the large scales (i.e., the decay rate exceeds the production rate). Conversely, during the accelerating phase ($\phi \in [\pi, 2\pi]$), $\langle F \rangle_{\phi}$ becomes negative, indicating a net storage of energy in the large scales. The amplitude of this oscillation serves as a direct, physical measure of the strength of the unsteadiness within the turbulent cascade. The DNSs present a similar behaviour (Fig.~\ref{fig:unsteady_function}(b)). We verified that the reported $F$ values are robust to the choice of evaluation scale, with qualitatively identical results obtained in the range $r \in [0.5\lambda, 2\lambda]$ (\hyperref[fig:S2]{End Matter, Fig.~6}). We also note that, given that $\langle F \rangle_{\phi}$ is defined in integral form in Eq.~(\ref{eq:unsteady_F}), it presents less scatter than the hysteresis cycles reported above.

Fig.~\ref{fig:unsteady_function}(c) shows the peak-to-peak amplitude of the unsteady function at the Taylor scale, $|\langle F(\lambda) \rangle_{\phi}|$, against the dimensionless forcing parameter, $X$. Remarkably, a good collapse is observed for all experimental and numerical data. Moreover, the relation $F(\lambda)\propto\sqrt{X}$ fits the data trend. This proves that our macroscopic forcing parameter is not only associated with the hysteresis area, but is a direct proxy for the magnitude of the underlying physical cause (the unsteady term in the KH equation). In consequence, the balance within the inertial range between the dissipative and the inter-scale energy transfer terms from Eq.~(\ref{eq:karman_howarth_main}) is altered by the effect of the unsteady term. This term can be both positive and negative during the cycle, producing a measurable effect in the dissipation constant $C_\varepsilon$. Moreover, when a solenoidal forcing is considered, as $F$ is defined positive for decelerating flows, the KH equation also justifies that the upper branch of the hysteresis cycle corresponds to such regime of the flow. This provides, therefore, a quantitative explanation of the gallery of hysteresis cycles shown in \hyperref[fig:S1]{the End Matter (Fig.~5)}, relating the intensity of the hysteresis cycle with the unsteady flow nature.

In summary, we have established a framework for the onset of hysteresis in the turbulent energy cascade. Combining HWA experiments and DNS of periodically forced turbulence, we identified a universal scaling law linking the hysteresis magnitude (measured by the normalized area of the $(\langle Re_{\lambda} \rangle_{\phi}, \langle C_{\epsilon} \rangle_{\phi})$ loop) to a single dimensionless forcing parameter, $X$. This law is physically grounded in the energy balance of the flow, as $X$ scales with the square of the unsteady term in the KH equation. This phenomenon constitutes a manifestation of \emph{dynamical hysteresis}~\cite{broner1997}: the loop area vanishes as $X \to 0$, as confirmed by our data, and no bistability or multiple attractors are required. Our findings thus reveal that hysteresis is a macroscopic signature of the time-dependent energy transfer within the cascade, providing a simple predictive link between large-scale forcing and instantaneous turbulent dissipation.% (i.e.\ as $f \to 0$ at fixed amplitude)

The implications of this finding are significant. It provides a simple predictive tool for estimating the degree to which a turbulent flow will depart from a quasi-steady state based on the characteristics of its unsteady boundary conditions. The ability to predict the magnitude of the cascade's hysteresis from simple, large-scale parameters is a crucial step towards developing more advanced models for a wide range of real-world out-of-equilibrium systems. This includes industrial processes that involve pulsating jets or mixers, and natural phenomena such as atmospheric turbulence interacting with gusts or the transport of pollutants and aerosols in unsteady winds, where understanding the unsteady dynamics is of paramount importance \cite{frisch1995turbulence, zapata2024turbulence}. \NEW{Time-dependent Navier--Stokes has two independent control parameters once the forcing is parametrised by $X$: at large $Re_\lambda$ the effects survive if $f$ is small enough that $St_f\sim\mathcal{O}(1)$, e.g.\ atmospheric flows ($L\sim 10^{3}$~m, $u'\sim 1$~m/s, $f\sim 10^{-3}$~Hz).} Crucially, while the present study employs periodic forcing as a controlled probe, the framework is not restricted to this case: since $C_\varepsilon$ at any instant reflects the forcing history, the response to arbitrary aperiodic signals can be reconstructed by decomposing the forcing into its Fourier components and applying the $\sqrt{X}$ scaling mode by mode. \NEW{More work is needed to generalize the framework to other flows and broader $X$ ranges, and to address the time lag between the Reynolds number and dissipation. A simple linear-response model with relaxation time $\tau_L = L/u'$ already predicts a phase lag $\Delta\phi \sim \arctan(St_f)$ consistent with the observed loop opening, but a rigorous derivation remains to be explored.}\\

\begin{acknowledgments}
JAF acknowledges support from Centrale Lille through a Guest Professor appointment in 2024 and for the experiments by the University of Colorado Boulder as part of his sabbatical. PDM acknowledges support from Centrale Lille through a Guest Professor appointment in 2025, and from proyecto REMATE of Redes Federales de Alto Impacto, Argentina. Part of this work has been performed during the 2024 and 2025 Lille Turbulence Programmes.
\end{acknowledgments}

%%============================================================
%% END MATTER
%%============================================================

\clearpage
\section*{End Matter}
\label{sec:endmatter}

%%------------------------------------------------------------
\subsection*{S1.~Experimental and DNS parameters}
\label{em:params}
%%------------------------------------------------------------

Table~\ref{tab:S1} lists all HWA and DNS conditions.
The dimensionless parameter $X = \mathrm{St}_f\cdot|\widetilde{U}|/\bar{U}$ (HWA) or $X = \mathrm{St}_f\cdot\delta A/A_0$ (DNS),
serves as the universal control variable.
The DNS $X$ values (0.021--0.084) overlap the upper HWA range (0.014--0.025),
enabling cross-validation of the $\sqrt{X}$ scaling across both datasets.

\begin{table}[h]
\caption{HWA experimental conditions (top) and DNS run parameters (bottom).
All HWA measurements at 3\,m downstream; $\bar{U}\approx15$\,m/s.
DNS: $N^3=512^3$, $\nu=5\times10^{-4}\,L_0U_0$,
steady-state $u'\approx0.54\,U_0$, $L\approx1.85\,L_0$,
$\mathrm{Re}_\lambda\approx150$, $C_\varepsilon\approx1.50$.}
\label{tab:S1}
\begin{ruledtabular}
\begin{tabular}{lcccccc}
Grid & $f$\,(Hz) & $|\widetilde{U}|/\bar{U}$ & $\mathrm{Re}_\lambda$ &
$C_\varepsilon$ & $\mathrm{St}_f$ & $X$ \\
\hline
FS & 0.05 & 0.20 & 196 & 0.47 & 0.0057  & 0.0011 \\
FS & 0.10 & 0.20 & 196 & 0.47 & 0.0102 & 0.0017 \\
FS & 0.25 & 0.20 & 196 & 0.47 & 0.0402  & 0.0040  \\
FS & 0.50 & 0.05 & 196 & 0.47 & 0.0342  & 0.0020  \\
FS & 0.50 & 0.10 & 196 & 0.47 & 0.0392  & 0.0059  \\
FS & 0.50 & 0.15 & 196 & 0.47 & 0.0378  & 0.0076  \\
FS & 0.50 & 0.20 & 196 & 0.47 & 0.0389   & 0.0078   \\
FS & 0.75 & 0.20 & 196 & 0.47 & 0.0552  & 0.0111  \\
FS & 1.00 & 0.20 & 196 & 0.47 & 0.0807  & 0.0161  \\
LG & 0.10 & 0.20 & 230 & 0.53 & 0.0125  & 0.00251  \\
LG & 0.50 & 0.20 & 230 & 0.53 & 0.0709  & 0.0142   \\
LG & 1.00 & 0.20 & 230 & 0.53 & 0.126   & 0.0252   \\
\hline
Run & $T\,(L_0/U_0)$ & $\delta A/A_0$ & $\mathrm{Re}_\lambda$ &
$C_\varepsilon$ & $\mathrm{St}_f$ & $X$ \\
\hline
\texttt{amp}  & 50   & 0.25 & 150 & 1.5 & 0.132 & 0.042 \\
\texttt{ampha} & 50 & 0.125 & 150 & 1.5 & 0.103 & 0.021 \\
\texttt{amphf}  & 25 & 0.25 & 150 & 1.5 & 0.328 & 0.084 \\
\end{tabular}
\end{ruledtabular}
\end{table}

%%------------------------------------------------------------
\subsection*{S2.~Velocity decomposition and phase averaging}
\label{em:decomp}
%%------------------------------------------------------------

The forcing frequency $f$ is identified from the power spectral density
peak, at least 30\,dB above the turbulent background.
The coherent component $\widetilde{U}(t)$ is extracted by a zero-phase
narrow-band IIR (Butterworth) filter centred at $f$ with a relative steepness of 0.95
(i.e.\ the $-3$\,dB point is at $0.95\,f$).
Phase bins of $\Delta\phi=20^\circ$ are formed using
$\phi(t)=2\pi f t+\phi_0$.
Statistics are averaged over 10 independent realisations $\approx600$ cycles at $f=0.5$\,Hz (HWA) or $\approx4$ forcing cycles (DNS).
Hysteresis is statistically significant ($>2\sigma$) for FS grid
$f\geq0.25$\,Hz with $|\widetilde{U}|/\bar{U}\geq0.10$, and for LG grid
$f\geq0.5$\,Hz; the quasi-static FS and LG cases (e.g. at $f=0.1$\,Hz) are marginally
significant ($1$--$2\sigma$), consistent with $X\to0$.

%%------------------------------------------------------------
\subsection*{S3.~K\'arm\'an--Howarth balance verification}
\label{em:khe}
%%------------------------------------------------------------

\begin{figure}
    \centering
    \includegraphics[width=0.95\linewidth, clip, trim=1.2cm 7.4cm 2.7cm 8.4cm]{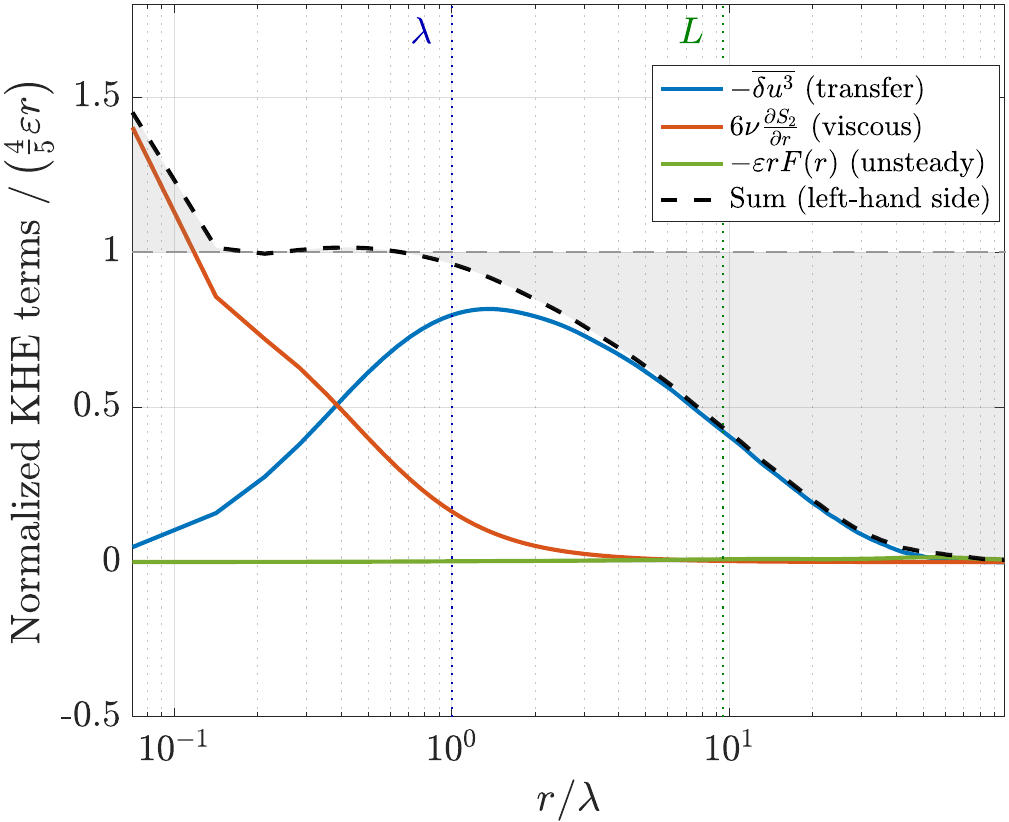}%, clip, trim=0.85cm 7.6cm 2.0cm 7.9cm
    % \fbox{\parbox[c][6cm][c]{0.49\textwidth}{\centering \textbf{Placeholder for Figure 3}}}
    
    \caption{Phase-averaged KH balance for
a representative case. The three left-hand terms of Eq.~\ref{eq:karman_howarth_main} are
computed and normalized by the turbulent dissipation term; the grey band shows the residual.}
\label{fig:S3}
\end{figure}

\begin{figure*}[t!]
\centering
\begin{subfigure}[b]{0.49\textwidth}
    \centering
    \includegraphics[width=0.95\linewidth,clip]{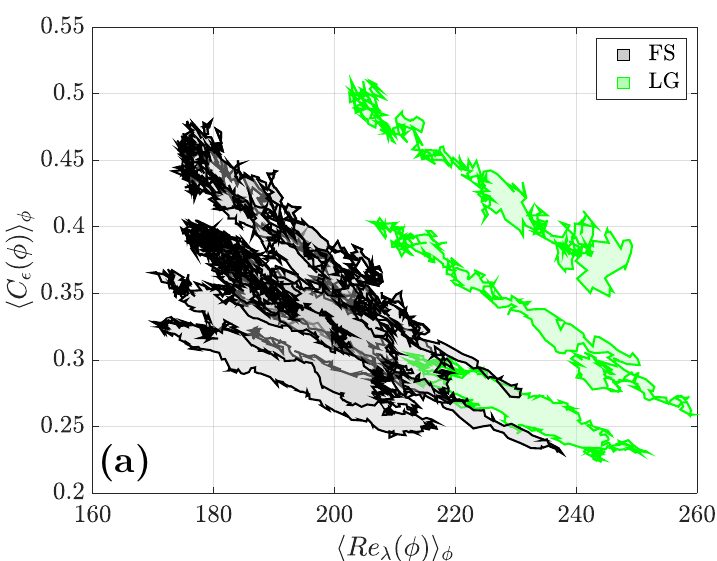}%,
        %trim=0.85cm 7.6cm 2.0cm 7.9cm
    \label{fig:S1a}
\end{subfigure}
\hfill
\begin{subfigure}[b]{0.49\textwidth}
    \centering
    \includegraphics[width=0.95\linewidth,clip]%
        {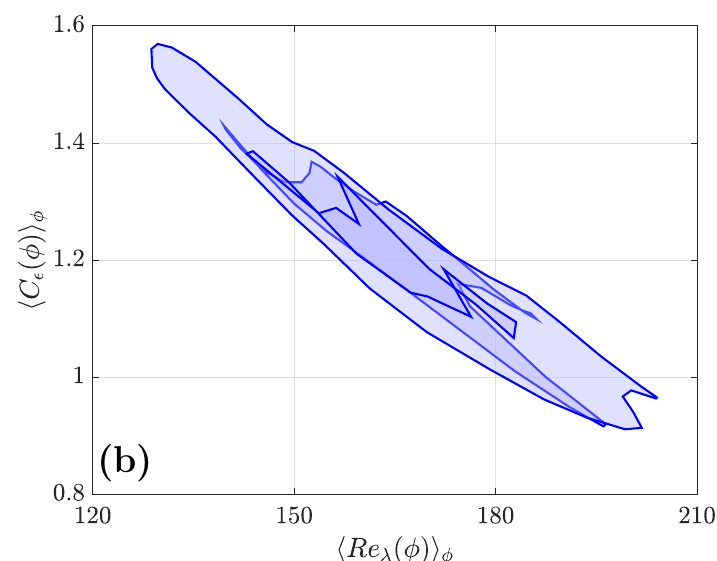}%,
        %trim=0.85cm 7.6cm 2.0cm 7.9cm
    \label{fig:S1b}
\end{subfigure}
\caption{Gallery of phase-averaged hysteresis loops in the
($\langle\mathrm{Re}_\lambda\rangle_\phi$, $\langle C_\varepsilon\rangle_\phi$)
plane for all conditions: (a)~HWA experiments (FS and LG grids, multiple
frequencies and amplitudes); (b)~DNS (\texttt{amp}, \texttt{ampha},
\texttt{amphf} runs).}%Shaded bands are standard deviations across 10 realisations (HWA) or 4 forcing cycles (DNS).
\label{fig:S1}
\end{figure*}

All four terms of Eq.~\ref{eq:karman_howarth_main} are computed phase-resolved.
For HWA: the inertial term from $S_3(r,\phi)$ via Taylor's hypothesis;
the viscous term from $\partial_r S_2$; the unsteady term from
$\partial_\phi\langle\delta u^2\rangle$ at known frequency; and
$\varepsilon(\phi)$ from the spectral method (integration of
$15\nu k^2 E(k)$, cross-checked via $15\nu\langle(\partial u/\partial x)^2\rangle$
to within 10--15\%).
The KHE residual is $\lesssim18\%$ of the dissipation term (HWA) and
$\approx30\%$ (DNS), where the higher value reflects finite-Reynolds-number
corrections and scale-by-scale injection/dissipation imbalance.
The phase-averaged budget for a representative case is shown in Fig.~\ref{fig:S3}.

% \begin{figure}[h]
% \centering
% \fbox{\parbox[c][5.5cm][c]{0.90\linewidth}{\centering
%   \textbf{Fig.~S3 — KHE phase-resolved budget}\\[6pt]
% }}
% \caption{Fig.~S3.\ Phase-resolved K\'arm\'an--Howarth balance for
% representative HWA and DNS cases. The four terms of Eq.~(1) are
% computed phase-resolved; the grey band shows the residual.}
% \label{fig:S3}
% \end{figure}

%%------------------------------------------------------------
\subsection*{S4.~Flow homogeneity}
\label{em:homog}
%%------------------------------------------------------------

At 3\,m downstream, prior measurements in the same facility~\cite{farnsworth2020design} confirmed mean-velocity uniformity within 2\%
across 80\% of the cross-section, turbulence-intensity uniformity within 5\%,
and bulk-velocity-oscillation spatial uniformity within 3\% ($f=1$\,Hz) to 1\%
($f=0.1$\,Hz).
The HWA probe is at the cross-section centre, where wall effects are minimal.

%%------------------------------------------------------------
%% Figure S1 -- Hysteresis gallery (former Fig.~2 of main text)
%%------------------------------------------------------------

%%------------------------------------------------------------
\subsection*{S5.~Scale sensitivity of $F(r)$}
\label{em:scale}
%%------------------------------------------------------------

\begin{figure}[h]
\centering
    \includegraphics[width=0.95\linewidth, clip]{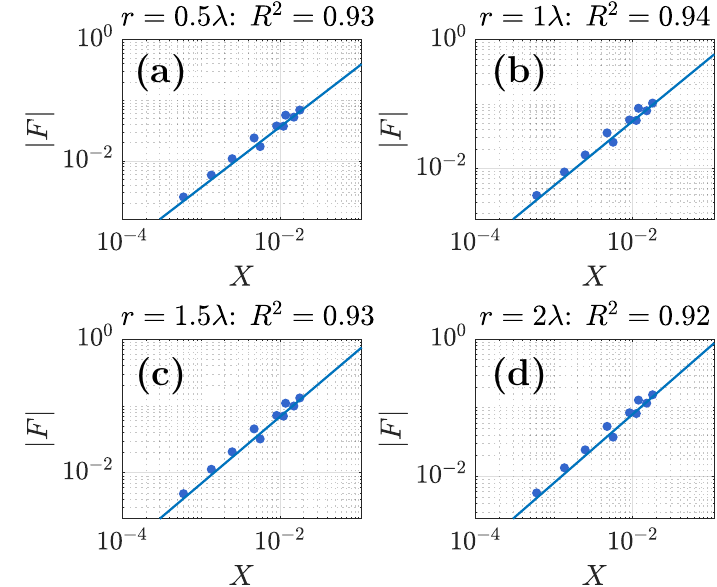}%, trim=1.2cm 7.9cm 2.7cm 7.8cm
% \fbox{\parbox[c][5cm][c]{0.90\linewidth}{\centering
%   \textbf{Fig.~S2 — Scale sensitivity of $F(r)$}\\[6pt]
% }}
\caption{Peak-to-peak amplitude $|F(r)|$ vs.\ $X$ at
$r = 0.5\lambda$, $\lambda$, $1.5\lambda$, and $2\lambda$ (a--d, respectively) for FS datasets. The scaling and collapse are preserved across all scales.}
\label{fig:S2}
\end{figure}

Figure~\ref{fig:S2} shows $|F(r)|$ vs.\ $X$ at $r=0.5\lambda$, $\lambda$,
$1.5\lambda$, and $2\lambda$.
The $\sqrt{X}$ scaling is preserved at all scales: best-fit exponent $b$
is $0.48\pm0.05$, $0.50\pm0.03$, $0.51\pm0.04$, and $0.52\pm0.05$,
respectively, all consistent with $b=1/2$, and the $R^2$ value at $r = L$ is 0.89.
The tightest cross-configuration collapse occurs at $r=\lambda$, supporting
the choice of the Taylor microscale.
This scale is also physically preferred because it represents, following
Lundgren's matched asymptotic analysis~\cite{lundgren2002,lundgren2003},
the point at which decaying turbulence sits closest to Kolmogorov
equilibrium~\cite{obligado2019quantifying,meldi2021analysis}.

%%------------------------------------------------------------
\subsection*{S6.~Necessity of the amplitude factor in $X$}
\label{em:necessity}
%%------------------------------------------------------------

\begin{figure}[h]
    \centering
    \includegraphics[width=0.95\linewidth, clip]{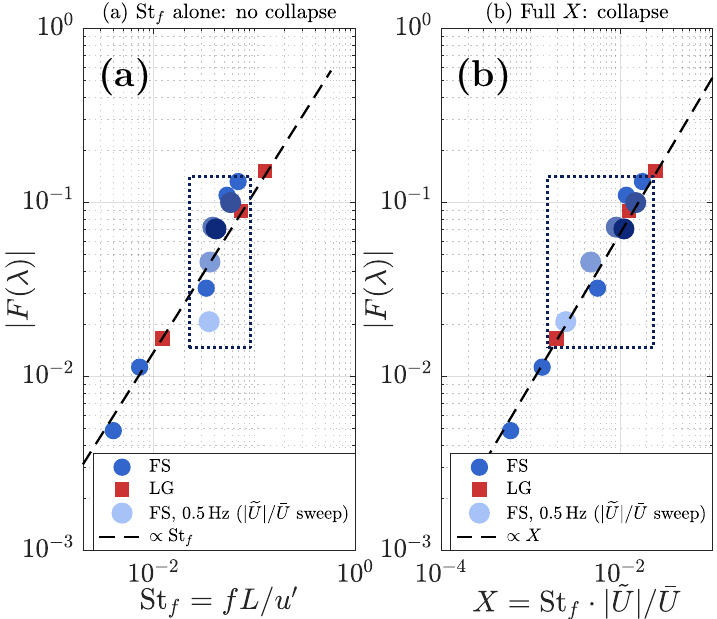}%, trim=1.5cm 7.4cm 2.7cm 7.8cm
\caption{Comparison of $|F(\lambda)|$ vs.\ (a) $\mathrm{St}_f$ alone
and (b) the full parameter $X$. The amplitude factor is necessary for data
collapse.}
\label{fig:S4}
\end{figure}

When $|F(\lambda)|$ is plotted against $\mathrm{St}_f$ alone, the data do
\emph{not} collapse: FS-grid cases at fixed $f=0.5$\,Hz span a factor of
$\approx5$ in $|F(\lambda)|$ across amplitudes
$|\widetilde{U}|/\bar{U}=0.05$--$0.20$, and DNS runs \texttt{amp} and
\texttt{ampha} (identical $\mathrm{St}_f$, half-amplitude difference) fall
at clearly different $|F(\lambda)|$ values.
Including the amplitude factor collapses all data onto the single $\sqrt{X}$
curve (Fig.~\ref{fig:S4}), unambiguously demonstrating that the amplitude
is an essential component of the scaling variable.

% \begin{figure}[h]
% \centering
% \fbox{\parbox[c][5cm][c]{0.90\linewidth}{\centering
%   \textbf{Fig.~S4 — Necessity of the amplitude factor in $X$}\\[6pt]
% }}
% \caption{Fig.~S4.\ Comparison of $|F(\lambda)|$ vs.\ (left) $\mathrm{St}_f$ alone
% and (right) the full parameter $X$. The amplitude factor is necessary for data
% collapse.}
% \label{fig:S4}
% \end{figure}

%%------------------------------------------------------------
\subsection*{S7.~Phase-lag analysis}
\label{em:phaselag}
%%------------------------------------------------------------

The phase lag visible in Fig.~\ref{fig:concepts}(b), where the
accelerating-to-decelerating transition does not occur at $\phi=0$, is a
direct signature of the cascade's finite relaxation time.
Modelling $\varepsilon(t)$ as a first-order linear relaxation toward the
quasi-static value $\varepsilon_\mathrm{qs}(t)\propto u'(t)^3$ with
relaxation time $\tau_L=L/u'$:
\begin{equation}
\frac{d\varepsilon}{dt} + \frac{\varepsilon}{\tau_L}
= \frac{\varepsilon_\mathrm{qs}(t)}{\tau_L},
\label{eq:relax_em}
\end{equation}
with sinusoidal forcing
$\varepsilon_\mathrm{qs}(t)=\bar\varepsilon(1+a\sin\omega t)$,
the steady-state solution yields a phase lag
\begin{equation}
\Delta\phi = \arctan(\mathrm{St}_f),
\label{eq:lag_em}
\end{equation}
where $\mathrm{St}_f = f\tau_L$.
For $\mathrm{St}_f\ll1$ ($X\to0$), $\Delta\phi\to0$ and the loop closes;
for $\mathrm{St}_f\sim\mathcal{O}(1)$, $\Delta\phi$ is tens of degrees,
opening the loop.
The same model predicts that the dissipation oscillation amplitude is
attenuated by $H(\mathrm{St}_f)=(1+\mathrm{St}_f^2)^{-1/2}$ relative to
the quasi-static value, consistent with the sub-linear $F(\lambda)\propto\sqrt{X}$
scaling in Fig.~\ref{fig:unsteady_function}(c).
A rigorous derivation of the exponent $1/2$ directly from the
KH equation remains an open problem.

\bibliographystyle{apsrev4-2}

\end{document}